%% file: 24ICASSP_LoFi-v4_final.tex
\documentclass{article}
\usepackage{amsmath,graphicx,./style/spconf}
\usepackage{cite}
\usepackage[T1]{fontenc}
\usepackage{amsmath, amsbsy}
\usepackage{amsthm}
\usepackage{subfigure}
\usepackage{booktabs}
\usepackage{multirow}
\usepackage{microtype}
\usepackage{balance}
\usepackage{colortbl}
\usepackage{xcolor}
\usepackage{bbm}
\usepackage[hidelinks]{hyperref}
\usepackage{comment}
\usepackage{circledsteps}

\input{macros/vmr-symbols-vecbold}
\input{macros/standard-macros}

\input{macros/defs}

\definecolor{mycolor7}{rgb}{0.63500,0.07800,0.18400}%

\newcommand{\PC}[1]{\ensuremath{\!\left(#1\right)}}

\title{LoFi USER SCHEDULING FOR MULTIUSER MIMO WIRELESS SYSTEMS}
\name{\begin{minipage}{0.95\textwidth}
\centering
\it Alexandra Gallyas-Sanhueza$^{\,\star1}$, Gian Marti$^{\,\star2}$, Victoria Palhares$^{\,\star2}$, \\ \it Reinhard Wiesmayr$^{\,\star2}$, and Christoph Studer$^2$\thanks{Simulator available at https://github.com/IIP-Group/lofi-user-scheduling.}
\end{minipage}}\address{\small $^\star$equal contribution and in alphabetic order; $^1$Cornell University, $^2$ETH Zurich \\ 
\small e-mail: ag753@cornell.edu, marti@iis.ee.ethz.ch, palhares@iis.ee.ethz.ch, wiesmayr@iis.ee.ethz.ch, studer@ethz.ch}

\begin{document}
\maketitle
\begin{abstract}
We propose new low-fidelity (LoFi) user equipment (UE) scheduling algorithms for multiuser multiple-input multiple-output (MIMO) wireless communication systems. The proposed methods rely on an efficient guess-and-check procedure that, given an objective function, performs paired comparisons between random subsets of UEs that should be scheduled in certain time slots. The proposed LoFi scheduling methods are computationally efficient, highly parallelizable, and gradient-free, which enables the use of almost arbitrary, non-differentiable objective functions. System simulations in a millimeter-wave (mmWave) multiuser MIMO scenario demonstrate that the proposed LoFi schedulers outperform a range of state-of-the-art user scheduling algorithms in terms of bit error-rate and/or computational complexity. 
\end{abstract}

\begin{keywords}
Linear MMSE, multiuser MIMO, paired comparisons, post-equalization SINR, UE scheduling.
\end{keywords}

\section{Introduction}
\label{sec:intro}

Next-generation wireless systems are expected to serve an increasing number of user equipments (UEs) in the same time-frequency resources with high data-rate requirements \cite{Ericsson}. One of the key challenges of such multiuser (MU) systems is managing inter-UE interference, which is typically caused either by very strong UEs or UEs that are nearby in space and suffer from correlated propagation conditions. 
In order to mitigate inter-UE interference, resource allocation strategies come to the rescue. 
Two common resource allocation mechanisms are power control and UE scheduling. Power control deals with the issue of some UEs causing significantly stronger receive signals at the infrastructure base station (BS) than others~\cite{Chiang2008}. 
In this work, we focus on UE scheduling, which distributes UE requests either over time, frequency, or space in order to improve quality-of-service (QoS). 
Specifically, we develop UE scheduling algorithms that assign certain UEs to certain time slots with the goal of optimizing a certain objective function that characterizes the desired QoS goals.

The literature on UE scheduling  mainly focuses on greedy algorithms \cite{Choi2019, Yoo2006, Lee2018a, Cui2018, He2017, Kim2018, Lee2018b, Gao2020, Paul2020, Wu2017, Zhu2021, Zhao2018, Xu2018, Zhang2022} which pick one UE at a time to join a set of scheduled UEs based on a certain objective function. 
Such algorithms, however, suffer from the limitation that once a scheduling set contains a problematic set of UEs, the performance remains poor. Furthermore, greedy procedures require a frequent re-evaluation of an objective function, which can be computationally expensive. 
Alternative scheduling algorithms are based on gradient descent \cite{Palhares2022} and self-supervised learning strategies, such as k-means clustering \cite{Chukhno2021,Zou2021}. Such scheduling algorithms consider the scheduling problem globally and often outperform greedy methods, but at the cost of increased computational complexity. 
The recent invasion of large language models (LLMs) into the realm of wireless communications \cite{du2023power} may soon transform the UE scheduling landscape as well, but we expect such methods to result in prohibitively high computational complexity.

\subsection{Contributions}

We propose novel, low-complexity UE scheduling algorithms that partition the UE set into separate time slots to minimize inter-UE interference.
The proposed methods, referred to as low-fidelity (LoFi) scheduling and LoFi++, generate random time slot assignments and perform paired comparisons for a given objective function to decide for the better schedule. 
In LoFi, we improve performance by leveraging randomized restarts; in LoFi++, we improve performance by including the opportunity to swap problematic UEs that are experiencing the highest interference in each time slot.
We evaluate LoFi and LoFi++ in terms of uncoded bit error-rate (BER) and complexity in a mmWave multiuser MIMO scenario with a worst-case signal-to-interference-plus-noise ratio (SINR) objective function and using channel vectors from a commercial ray-tracer \cite{remcom}. 
To demonstrate the efficacy of our approach, we compare our method to multiple baselines, including an exhaustive search.

\subsection{Notation}
Upper-case and lower-case bold symbols represent matrices and vectors, respectively. Upper-case calligraphic letters represent sets.
We use $\bma_i$ for the $i$th column of $\bA$ and $\bA_{\setA}$ for the submatrix of $\bA$ consisting of the columns indexed by the set~$\setA$. The cardinality of the set $\setA$ is $|\setA|$.
The $N \times N$ identity matrix is~$\bI_N$.  The Euclidean norm of a vector $\bma$ is $\vecnorm{\bma}_2$. The superscript~${}\PC{\cdot}^H$ represents Hermitian transposition.

\section{System Model}

\begin{figure}[tp]
\centering
\includegraphics[width=0.99\columnwidth]{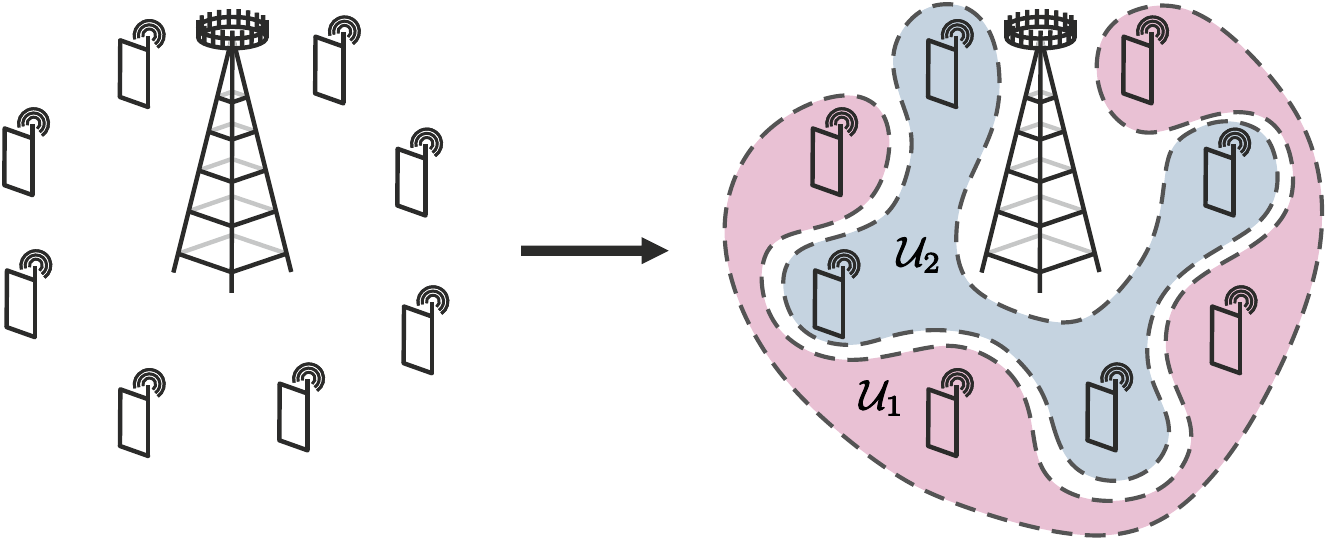}
\vspace{-0.1cm}
\caption{Example of a set $\setU$ of UEs that is scheduled into two disjoint groups $\setU_1$ and~$\setU_2$.}
\label{fig:scheduling_diagram}
\end{figure}

We consider the uplink of a MU-MIMO system in which~$U$ single-antenna UEs transmit data to a $B$-antenna BS.
The channel vector of the $u$th UE is denoted by $\bmh_u\in\opC^B$, and $\bH=[\bmh_1,\dots,\bmh_U] \in \opC^{B \times U}$ is the channel matrix describing the propagation conditions between all of the UEs and the BS. 
Determining a schedule for two time slots amounts to identifying a partition of the set of UE indices $\setU=\{1,2,\ldots,U\}$ into two groups $\firstslotset$ and $\secondslotset$ of equal cardinality $|\firstslotset|=|\secondslotset|$ while optimizing a given objective.\footnote{By a partition, we mean that $\firstslotset\cap\secondslotset=\{\}$ and  $\firstslotset\cup\secondslotset=\setU$.}
We call such a partition a \emph{schedule} and denote it as $\randomset = (\firstslotset, \secondslotset)$; see \fref{fig:scheduling_diagram}.
Note that any such schedule is intrinsically fair since any UE gets scheduled for the same relative duration (once every other time slot).
The input-output relationship per time slot can be~written~as 
\begin{align}
    \bmy_i = \bH_{\setU_i}\bms_{\setU_i} + \bmn_i, \,\, i\in\{1,2\}.
\end{align}
Here,  $\bmy_i\in\opC^{B}$ is the BS receive vector in time slot $i$, 
$\bH_{\setU_i}$ is the submatrix of $\bH$ consisting of the columns indexed by $\setU_i$, 
$\bms_{\setU_i}\in\setQ^{U/2}$ is the transmit vector by the UEs scheduled in time slot $i$ with elements taken from a constellation
$\setQ$ with average symbol energy $\Es$, and $\bmn_i$ is i.i.d.\ circularly-symmetric complex Gaussian noise with variance $\No$. 

Both for computing a schedule and for detecting the transmitted data, we assume that the BS obtains estimates $\hat\bmh_u$ of all channel vectors $\bmh_u$ using BEACHES~\cite{Mirfarshbafan2020}.
The BS uses linear minimum mean squared error (LMMSE) data detection for each time slot, i.e., 
the estimate $\hat\bms_{\setU_i}$ of $\bms_{\setU_i}$ is obtained by 
\begin{align}
    \hat\bms_{\setU_i} = \herm{\bW_{\setU_i}}\bmy_i, 
\end{align}
where $\bW_{\setU_i}=\hat\bH_{\setU_i}\inv{(\herm{\hat\bH_{\setU_i}}\hat\bH_{\setU_i}+\frac{\No}{\Es}\bI_{U/2})}$
is the LMMSE equalization matrix, with $\hat\bH_{\setU_i}$ denoting the estimated channel matrix of the scheduled UEs. 
In what follows, our goal of scheduling is to achieve the lowest BER, averaged over all~UEs. 

\section{L\lowercase{o}F\lowercase{i} User Scheduling Algorithms} 

The first insight for our approach is that the BER at high signal-to-noise ratio (SNR) is dominated by the worst UE. 
Our goal is therefore to maximize the post-equalization SINR of the worst UE \cite{Ngo2017}.\footnote{While we use this as a proxy for the BER, maximizing the performance of the worst
UE could also be seen as a legitimate objective in its own right.}
For a given schedule~$\setS$, the post-equalization SINR of the $u$th UE is given by
\begin{align}
    \textit{SINR}_u(\randomset) = 
    \frac{|\herm{\bmw_u}\hat\bmh_u|^2}{\sum_{u'\in\setU_\ast(u), u'\neq u}|\herm{\bmw_u}\hat\bmh_{u'}|^2
    + \frac{\No}{\Es}\|\bmw_u\|_2^2},
\end{align}
with $\bmw_u$ being the column of ${\bW_{\setU_\ast(u)}}$ which corresponds to the $u$th UE, where $\setU_\ast(u)=\firstslotset$ if $u\in\firstslotset$ and $\setU_\ast(u)=\secondslotset$ otherwise.
Our objective function is therefore 
\begin{align} \label{eq:objective}
    \max_{\setS}\, \min_{u\in\setU}\, \textit{SINR}_u(\randomset),
\end{align}
which aims at maximizing the worst per-UE post-equalization SINR. 
Note that this objective function is not differentiable, not just because of the discrete nature of the partition $\setS$, 
but also because of the minimum over the discrete set~$\setU$. Thus, gradient-based scheduling methods, such as the one proposed in~\cite{Palhares2022}, cannot be used.  
As we will show next, the gradient-free nature of the proposed LoFi and LoFi++ methods are able to deal with this limitation in a computationally efficient manner.\footnote{
Note that, while LoFi and LoFi++ use the worst-case post-equalization SINR as objective, other objectives
(e.g. the average SINR, or the achievable sum rate) would also be possible.
} 

\subsection{LoFi: Guess-and-Check}

The second insight is that, while finding an optimal schedule is generally difficult, avoiding a terrible one is easy. 
In fact, a \emph{random} schedule performs often much closer to optimal schedule than avoiding scheduling entirely~\cite{Palhares2022}.
Of course, whether a randomly selected schedule performs well or poorly is a question of pure luck. 
But the chances of obtaining an \emph{extremely} bad schedule can be substantially reduced if one 
simply draws \emph{two} random schedules. 
The idea of LoFi is therefore to draw two random schedules, $\setS$ and $\setS'$, followed by 
computing $\min_{u\in\setU}\textit{SINR}_u(\setS)$ and $\min_{u\in\setU}\textit{SINR}_u(\setS')$, 
and deploying the schedule with the larger objective value. 

The computational cost amounts to evaluating $\textit{SINR}_u$ for all $u\in\setU$ for 
both schedules $\setS$ and $\setS'$ and is dominated by the cost of computing the four 
equalization matrices $\bW_{\setU_1}$, $\bW_{\setU_2}$, $\bW_{\setU_1'}$, and $\bW_{\setU_2'}$. 
Note, however, that the two equalization matrices associated with the deployed schedule would 
have to be calculated anyway for data detection. Thus, the computational overhead of scheduling itself is marginal.
Note also that the objectives of both schedules can be evaluated in parallel.

\subsection{LoFi++: Beyond Random Guessing}
\label{sec:lofiplusplus}

\begin{figure}[tp]
\centering
\includegraphics[width=0.55\columnwidth]{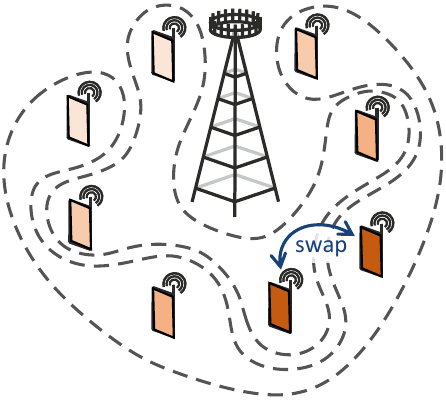}
\vspace{-0.1cm}
\caption{LoFi++ tests whether swapping the worst UE of each group is beneficial and swaps those UEs if so.
Darker UE shades indicate lower post-equalization SINR.}
\label{fig:lofi_plusplus}
\end{figure}

LoFi already significantly reduces the probability of obtaining a bad schedule, 
and therefore improves the average BER (see \fref{sec:simulation_results}), 
but it still purely depends on luck. 
Our improved variant, LoFi++, takes a more deliberate approach to avoid selecting a bad schedule, 
without increasing computational complexity. 
LoFi++ starts by drawing \emph{one} random schedule~$\setS$. 
It then determines the worst UE in both subgroups, i.e., it finds
\begin{align}
    \!\check{u}_1=\argmin_{u\in\firstslotset}\textit{SINR}_u(\electionone{\randomset}) 
    \,\, \text{and} \,\,
    \check{u}_2=\argmin_{u\in\secondslotset}\textit{SINR}_u(\electionone{\randomset}).
\end{align}
Instead of drawing another random schedule $\setS'$ and hoping for the best, 
LoFi++  actively tries to improve the SINRs of $\check{u}_1$ and $\check{u}_2$. 
It is clear that, of all UEs, $\check{u}_1$ and $\check{u}_2$ are those that are most disadvantaged
by being included in their respective subgroups, which implies that they are heavily affected by 
interference caused by other UEs in their subgroup. 
But there is no reason to expect that they would 
be as heavily affected by inter-UE interference if they were in the \emph{other} subgroup. 
Thus, instead of drawing a second schedule $\setS'=(\electiontwo{\firstslotset}, \electiontwo{\secondslotset})$ at random, 
LoFi++ creates a second schedule by swapping the users $\check{u}_1$ and~$\check{u}_2$ (see \fref{fig:lofi_plusplus} for an illustration):
\begin{align}
    \!\!\!\electiontwo{\firstslotset} = (\firstslotset \setminus \{\check{u}_1\})\cup\{\check{u}_2\}
    \,\, \text{and} \,\,
    \electiontwo{\secondslotset} = (\secondslotset \setminus \{\check{u}_2\})\cup\{\check{u}_1\}.\!\!\!
\end{align}
LoFi++ then compares the objectives $\min_{u\in\setU} \textit{SINR}_u(\randomset)$
and $\min_{u\in\setU} \textit{SINR}_u(\electiontwo{\randomset})$, 
and deploys the schedule with the higher objective value.
The computational complexity
of LoFi++ is exactly the same as that of LoFi. In contrast to LoFi, however, 
LoFi++ cannot parallelize evaluation of both schedules since $\setS'$
can only be computed \emph{after} $\setS$ has been evaluated.

\subsection{Improving LoFi/LoFi++ Through Random Restarts}

In the basic form, LoFi simply compares two random schedules and picks the better one.
Evidently, the performance can be improved further if one uses $K>1$ random restarts 
(i.e., one draws a total of $2K$ random schedules) and deploys the best of all drawn schedules.
The computational cost of such a random restarting procedure increases linearly with $K$. 

While LoFi++ as described in \fref{sec:lofiplusplus} compares a random schedule with a deterministic derivative schedule (instead of two independent random schedules), its chances of finding a good schedule can also be improved with $K$ random restarts and picking the best of all $2K$ considered schedules. 

Note that LoFi with $K$ random restarts and LoFi++ with~$K$ random restarts require evaluating the same number
of schedules (at the same computational cost for evaluating each schedule) and, thus, have the same computational complexity. However, LoFi can completely parallelize all of the $2K$ evaluations, while LoFi++ can parallelize at most 
$K$ of them, since the second half of candidate schedules can only be computed \emph{after} the initial 
candidate schedules have been evaluated.

\section{Simulation Results}
\label{sec:simulation_results}

\subsection{Simulation Setup}

In our simulations, we replicate the scenario from \cite{Palhares2022}. We consider the uplink of a MU-MIMO system with a BS with $B=16$ antennas deployed in a uniform linear array (ULA) with half-wavelength antenna spacing, and $U=16$ single-antenna UEs. 
The BS and the UEs are placed at a height of $10$\,m and $1.65$\,m, respectively. Both on the BS and UE side, we consider omnidirectional antennas. The carrier frequency is $60$\,GHz with a bandwidth of $100$\,MHz.
The transmit constellation $\setQ$ is 16-QAM, and the UEs use power control with a BS-side dynamic range of $6$~dB.
Our simulations use mmWave channel vectors obtained from the Wireless Insite ray-tracing software \cite{remcom}. 
In the simulated outdoor scenario, we consider $22\,448$ possible UE positions in a $109.7$\,m $\times$ $164.7$\,m area. 
For every channel realization, we pick $U$ UE positions uniformly and independently at random. 
We average the performance over $100$ channel realizations, and for each channel realization, we compute the uncoded BER over $10^5$ symbol transmissions.

\subsection{Baseline Algorithms and Performance Metrics}

To showcase the performance of LoFi and LoFi++, we compare them against multiple baselines. 
As baselines, we consider the greedy methods ``SUS'' \cite{Yoo2006}, ``CSS'' \cite{Choi2019}, and ``greedy'' \cite{Choi2019}. We also compare against the gradient-based method ``opt.-based'' from \cite{Palhares2022}, which considers the post-equalization mean squared error (MSE) as its objective. We also include the performance of an exhaustive search over all ${U\choose U/2}=12\,870$ possible schedules, where the quality of
a schedule is also based on the post-equalization MSE.
Finally, we show the performance of purely random scheduling as well as no scheduling (i.e., serving every UE in every time slot). 
Note that (apart from random scheduling and no scheduling), the computational complexity of all of these methods exceeds the complexity of LoFi and LoFi++; in the case of ``opt.-based'' and exhaustive search by orders of magnitude.\footnote{The number of times the greedy methods must evaluate their objectives is linear in the number of scheduled UEs. In contrast, the number of objective function evaluations is typically a very small constant for LoFi and LoFi++.}
As a performance metric, we use the uncoded BER averaged over all UEs.

\subsection{Simulation Results}

\begin{figure}[tb]
\centering
\includegraphics[width=0.85\linewidth]{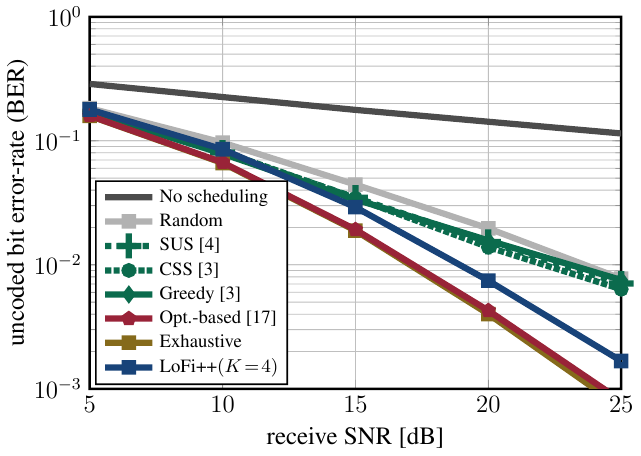} 
\caption{%
BER of the proposed LoFi++ method compared to state-of-the-art baselines. LoFi++ outperforms most of the baselines, except the ``opt.-based'' and exhaustive search, both of which require significantly higher computational complexity.}
\label{fig:ber_vs_snr_plot}
\end{figure}

In \fref{fig:ber_vs_snr_plot}, we see that the proposed LoFi++ method with $K=4$ random restarts
reaches a BER of approximately $0.2\%$ at an SNR of $25$\,dB. Compared to ``SUS,'' ``CSS,'' ``greedy,'' and random scheduling, 
LoFi++ achieves an improvement of at least $3$\,dB at $1\%$ BER.
While LoFi++ does not match the performance of the ``opt.-based'' baseline from \cite{Palhares2022} (or that of an exhaustive search), its complexity is orders of magnitude lower than the complexity of these two methods.
See \fref{tbl:runtime} for the runtime comparison between LoFi(++) and baselines, where all runtimes are normalized relative to the fastest method which is random scheduling.
We also highlight the importance of scheduling by noting the significant performance gains of \emph{any} kind of scheduling strategy---even the random schedule---compared to no scheduling that serves all users in all time~slots. 

\begin{figure}[tb]
\centering
\includegraphics[width=0.85\linewidth]{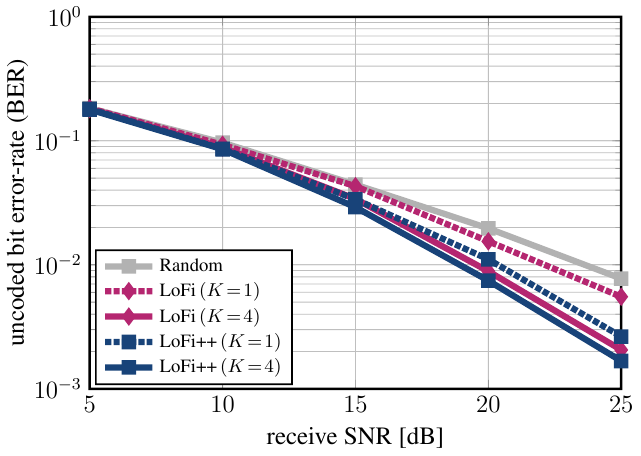} 
\caption{Comparison between LoFi and LoFi++. Increasing the number of random restarts $K$ improves the BER. }
\label{fig:ber_vs_snr_plot_compare_lofi}
\end{figure}

In \fref{fig:ber_vs_snr_plot_compare_lofi}, we compare the BER performance of LoFi and LoFi++ for different numbers $K$ of random restarts.
As expected, the performance of both methods increases as we increase $K$. 
In addition, LoFi++ always outperforms LoFi (for the same number of random restarts, which corresponds to the same computational complexity). This shows that, in addition to the diversity brought by restarting $K$ times, actively trying to improve the worst UEs of each scheduled group simply by swapping them works better than na\"ively drawing another random schedule.

\begin{table}[tbp]
\centering
\caption{Normalized MATLAB runtime comparison between LoFi(++) and baselines (``Random'' has normalized runtime~1). \vspace{.01cm}}
\label{tbl:runtime}	
\renewcommand{\arraystretch}{1.05}
\small	
\begin{tabular}{lrclr}
\toprule
\bf Method & \bf Runtime& & \bf Method & \bf Runtime \\ 
\midrule
Random & $\hphantom{0}1\times$ & ~~~~ & Opt.-based [17] & $236162\times$\\
SUS [4] & $\hphantom{0}5.5\times$ & & Exhaustive & $\hphantom{0}9808\times$\\
CSS [3]& $\hphantom{0}50\times$ & & LoFi (K=4) & $\hphantom{0}10\times$\\
Greedy [3]& $\hphantom{0}32\times$ & & LoFi++ (K=4) & $\hphantom{0}9\times$\\
\bottomrule 
\end{tabular}		
\end{table}

\begin{figure}[tb]
\centering
\hspace{2pt}
\includegraphics[width=0.8\linewidth]{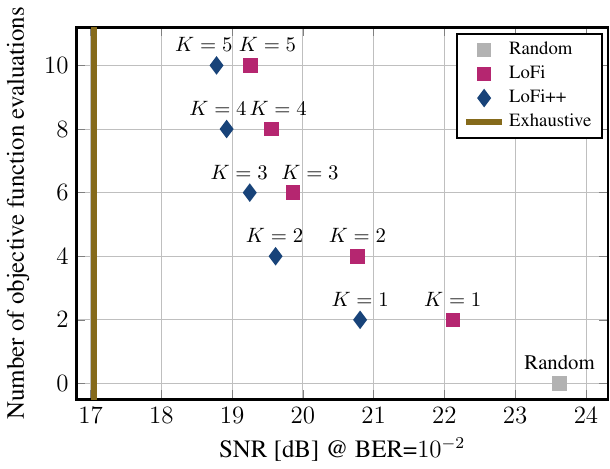}
\hspace{2pt}
\caption{Complexity versus SNR performance at 1\% BER for the proposed LoFi and LoFi++ UE scheduling methods.}
\label{fig:complexity_plot}
\end{figure}

In \fref{fig:complexity_plot}, we plot the number of objective function evaluations versus the minimum SNR (in dB) required to achieve a target BER of 1\%.
The y-axis serves as a proxy to computational complexity, while the x-axis is a horizontal slice through the BER plot in \fref{fig:ber_vs_snr_plot_compare_lofi}.
The desired location in this trade-off plot is the lower left, i.e., we seek schedules that require the lowest complexity at the lowest SNR operating point. 
Increasing the number of random restarts $K$ increases the complexity linearly; at the same time, the SNR operation point reduces. Nonetheless, we observe diminishing returns by increasing $K$ from $4$ to $5$, which indicates that $K=4$ is a sensible choice in this scenario. \fref{fig:complexity_plot} also confirms our earlier observation: LoFi++ consistently outperforms LoFi. We reiterate, however, that LoFi exhibits better parallelization abilities than LoFi++.

\section{Conclusions}

We have proposed novel UE scheduling methods that enable the use of nearly arbitrary objective functions while being computationally efficient and highly parallelizable. Our simulation results for a mmWave MU-MIMO scenario have shown that the resulting algorithms, which we call LoFi and LoFi++, outperform three state-of-the-art schedulers from \cite{Choi2019,Yoo2006} in performance and \cite{Palhares2022} in complexity. The generality of our methods regarding the use of a non-differentiable objective function has great potential to further optimize the quality of service in MU wireless systems. An in-depth exploration of other objective functions is part of ongoing work.

\balance

\end{document}

%% file: macros/vmr-symbols-vecbold.tex
\usepackage{amssymb}
\usepackage{amsfonts}
\usepackage{mathrsfs}
\usepackage{xspace}
\usepackage{bm}
\usepackage{upgreek}

\newcommand{\safemath}[2]{\newcommand{#1}{\ensuremath{#2}\xspace}}

\safemath{\bma}{\mathbf{a}}
\safemath{\bmb}{\mathbf{b}}
\safemath{\bmc}{\mathbf{c}}
\safemath{\bmd}{\mathbf{d}}
\safemath{\bme}{\mathbf{e}}
\safemath{\bmf}{\mathbf{f}}
\safemath{\bmg}{\mathbf{g}}
\safemath{\bmh}{\mathbf{h}}
\safemath{\bmi}{\mathbf{i}}
\safemath{\bmj}{\mathbf{j}}
\safemath{\bmk}{\mathbf{k}}
\safemath{\bml}{\mathbf{l}}
\safemath{\bmm}{\mathbf{m}}
\safemath{\bmn}{\mathbf{n}}
\safemath{\bmo}{\mathbf{o}}
\safemath{\bmp}{\mathbf{p}}
\safemath{\bmq}{\mathbf{q}}
\safemath{\bmr}{\mathbf{r}}
\safemath{\bms}{\mathbf{s}}
\safemath{\bmt}{\mathbf{t}}
\safemath{\bmu}{\mathbf{u}}
\safemath{\bmv}{\mathbf{v}}
\safemath{\bmw}{\mathbf{w}}
\safemath{\bmx}{\mathbf{x}}
\safemath{\bmy}{\mathbf{y}}
\safemath{\bmz}{\mathbf{z}}
\safemath{\bmzero}{\mathbf{0}}
\safemath{\bmone}{\mathbf{1}}
\safemath{\Bell}{\ensuremath{\boldsymbol\ell}}

\bmdefine{\biad}{a}
\bmdefine{\bibd}{b}
\bmdefine{\bicd}{c}
\bmdefine{\bidd}{d}
\bmdefine{\bied}{e}
\bmdefine{\bifd}{f}
\bmdefine{\bigd}{g}
\bmdefine{\bihd}{h}
\bmdefine{\biid}{i}
\bmdefine{\bijd}{j}
\bmdefine{\bikd}{k}
\bmdefine{\bild}{l}
\bmdefine{\bimd}{m}
\bmdefine{\bind}{n}
\bmdefine{\biod}{o}
\bmdefine{\bipd}{p}
\bmdefine{\biqd}{q}
\bmdefine{\bird}{r}
\bmdefine{\bisd}{s}
\bmdefine{\bitd}{t}
\bmdefine{\biud}{u}
\bmdefine{\bivd}{v}
\bmdefine{\biwd}{w}
\bmdefine{\bixd}{x}
\bmdefine{\biyd}{y}
\bmdefine{\bizd}{z}

\bmdefine{\bixid}{\xi}
\bmdefine{\bilambdad}{\lambda}
\bmdefine{\bimud}{\mu}
\bmdefine{\bithetad}{\theta}
\bmdefine{\biphid}{\phi}
\bmdefine{\bideltad}{\delta}

\safemath{\bmia}{\biad}
\safemath{\bmib}{\bibd}
\safemath{\bmic}{\bicd}
\safemath{\bmid}{\bidd}
\safemath{\bmie}{\bied}
\safemath{\bmif}{\bifd}
\safemath{\bmig}{\bigd}
\safemath{\bmih}{\bihd}
\safemath{\bmii}{\biid}
\safemath{\bmij}{\bijd}
\safemath{\bmik}{\bikd}
\safemath{\bmil}{\bild}
\safemath{\bmim}{\bimd}
\safemath{\bmin}{\bind}
\safemath{\bmio}{\biod}
\safemath{\bmip}{\bipd}
\safemath{\bmiq}{\biqd}
\safemath{\bmir}{\bird}
\safemath{\bmis}{\bisd}
\safemath{\bmit}{\bitd}
\safemath{\bmiu}{\biud}
\safemath{\bmiv}{\bivd}
\safemath{\bmiw}{\biwd}
\safemath{\bmix}{\bixd}
\safemath{\bmiy}{\biyd}
\safemath{\bmiz}{\bizd}

\safemath{\bmxi}{\bixid}
\safemath{\bmlambda}{\bilambdad}
\safemath{\bmmu}{\bimud}
\safemath{\bmtheta}{\bithetad}
\safemath{\bmphi}{\biphid}
\safemath{\bmdelta}{\bideltad}

\safemath{\bA}{\mathbf{A}}
\safemath{\bB}{\mathbf{B}}
\safemath{\bC}{\mathbf{C}}
\safemath{\bD}{\mathbf{D}}
\safemath{\bE}{\mathbf{E}}
\safemath{\bF}{\mathbf{F}}
\safemath{\bG}{\mathbf{G}}
\safemath{\bH}{\mathbf{H}}
\safemath{\bI}{\mathbf{I}}
\safemath{\bJ}{\mathbf{J}}
\safemath{\bK}{\mathbf{K}}
\safemath{\bL}{\mathbf{L}}
\safemath{\bM}{\mathbf{M}}
\safemath{\bN}{\mathbf{N}}
\safemath{\bO}{\mathbf{O}}
\safemath{\bP}{\mathbf{P}}
\safemath{\bQ}{\mathbf{Q}}
\safemath{\bR}{\mathbf{R}}
\safemath{\bS}{\mathbf{S}}
\safemath{\bT}{\mathbf{T}}
\safemath{\bU}{\mathbf{U}}
\safemath{\bV}{\mathbf{V}}
\safemath{\bW}{\mathbf{W}}
\safemath{\bX}{\mathbf{X}}
\safemath{\bY}{\mathbf{Y}}
\safemath{\bZ}{\mathbf{Z}}

\safemath{\bZero}{\mathbf{0}}
\safemath{\bOne}{\mathbf{1}}
\safemath{\bDelta}{\mathbf{\Delta}}
\safemath{\bLambda}{\mathbf{\UpLambda}}
\safemath{\bPhi}{\mathbf{\Upphi}}
\safemath{\bSigma}{\mathbf{\Upsigma}}
\safemath{\bOmega}{\mathbf{\Upomega}}
\safemath{\bTheta}{\mathbf{\Uptheta}}

\bmdefine{\biAd}{A}
\bmdefine{\biBd}{B}
\bmdefine{\biCd}{C}
\bmdefine{\biDd}{D}
\bmdefine{\biEd}{E}
\bmdefine{\biFd}{F}
\bmdefine{\biGd}{G}
\bmdefine{\biHd}{H}
\bmdefine{\biId}{I}
\bmdefine{\biJd}{J}
\bmdefine{\biKd}{K}
\bmdefine{\biLd}{L}
\bmdefine{\biMd}{M}
\bmdefine{\biOd}{N}
\bmdefine{\biPd}{O}
\bmdefine{\biQd}{P}
\bmdefine{\biRd}{R}
\bmdefine{\biSd}{S}
\bmdefine{\biTd}{T}
\bmdefine{\biUd}{U}
\bmdefine{\biVd}{V}
\bmdefine{\biWd}{W}
\bmdefine{\biXd}{X}
\bmdefine{\biYd}{Y}
\bmdefine{\biZd}{Z}

\bmdefine{\biDelta}{\Delta}
\bmdefine{\biLambda}{\Lambda}
\bmdefine{\biPhi}{\Phi}
\bmdefine{\biSigma}{\Sigma}
\bmdefine{\biOmega}{\Omega}
\bmdefine{\biTheta}{\Theta}

\safemath{\bimA}{\biAd}
\safemath{\bimB}{\biBd}
\safemath{\bimC}{\biCd}
\safemath{\bimD}{\biDd}
\safemath{\bimE}{\biEd}
\safemath{\bimF}{\biFd}
\safemath{\bimG}{\biGd}
\safemath{\bimH}{\biHd}
\safemath{\bimI}{\biId}
\safemath{\bimJ}{\biJd}
\safemath{\bimK}{\biKd}
\safemath{\bimL}{\biLd}
\safemath{\bimM}{\biMd}
\safemath{\bimN}{\biNd}
\safemath{\bimO}{\biOd}
\safemath{\bimP}{\biPd}
\safemath{\bimQ}{\biQd}
\safemath{\bimR}{\biRd}
\safemath{\bimS}{\biSd}
\safemath{\bimT}{\biTd}
\safemath{\bimU}{\biUd}
\safemath{\bimV}{\biVd}
\safemath{\bimW}{\biWd}
\safemath{\bimX}{\biXd}
\safemath{\bimY}{\biYd}
\safemath{\bimZ}{\biZd}

\safemath{\bimDelta}{\biDelta}
\safemath{\bimLambda}{\biLambda}
\safemath{\bimPhi}{\biPhi}
\safemath{\bimSigma}{\biSigma}
\safemath{\bimOmega}{\biOmega}
\safemath{\bimTheta}{\biTheta}

\safemath{\setA}{\mathcal{A}}
\safemath{\setB}{\mathcal{B}}
\safemath{\setC}{\mathcal{C}}
\safemath{\setD}{\mathcal{D}}
\safemath{\setE}{\mathcal{E}}
\safemath{\setF}{\mathcal{F}}
\safemath{\setG}{\mathcal{G}}
\safemath{\setH}{\mathcal{H}}
\safemath{\setI}{\mathcal{I}}
\safemath{\setJ}{\mathcal{J}}
\safemath{\setK}{\mathcal{K}}
\safemath{\setL}{\mathcal{L}}
\safemath{\setM}{\mathcal{M}}
\safemath{\setN}{\mathcal{N}}
\safemath{\setO}{\mathcal{O}}
\safemath{\setP}{\mathcal{P}}
\safemath{\setQ}{\mathcal{Q}}
\safemath{\setR}{\mathcal{R}}
\safemath{\setS}{\mathcal{S}}
\safemath{\setT}{\mathcal{T}}
\safemath{\setU}{\mathcal{U}}
\safemath{\setV}{\mathcal{V}}
\safemath{\setW}{\mathcal{W}}
\safemath{\setX}{\mathcal{X}}
\safemath{\setY}{\mathcal{Y}}
\safemath{\setZ}{\mathcal{Z}}
\safemath{\emptySet}{\varnothing}

\safemath{\colA}{\mathscr{A}}
\safemath{\colB}{\mathscr{B}}
\safemath{\colC}{\mathscr{C}}
\safemath{\colD}{\mathscr{D}}
\safemath{\colE}{\mathscr{E}}
\safemath{\colF}{\mathscr{F}}
\safemath{\colG}{\mathscr{G}}
\safemath{\colH}{\mathscr{H}}
\safemath{\colI}{\mathscr{I}}
\safemath{\colJ}{\mathscr{J}}
\safemath{\colK}{\mathscr{K}}
\safemath{\colL}{\mathscr{L}}
\safemath{\colM}{\mathscr{M}}
\safemath{\colN}{\mathscr{N}}
\safemath{\colO}{\mathscr{O}}
\safemath{\colP}{\mathscr{P}}
\safemath{\colQ}{\mathscr{Q}}
\safemath{\colR}{\mathscr{R}}
\safemath{\colS}{\mathscr{S}}
\safemath{\colT}{\mathscr{T}}
\safemath{\colU}{\mathscr{U}}
\safemath{\colV}{\mathscr{V}}
\safemath{\colW}{\mathscr{W}}
\safemath{\colX}{\mathscr{X}}
\safemath{\colY}{\mathscr{Y}}
\safemath{\colZ}{\mathscr{Z}}

\safemath{\opA}{\mathbb{A}}
\safemath{\opB}{\mathbb{B}}
\safemath{\opC}{\mathbb{C}}
\safemath{\opD}{\mathbb{D}}
\safemath{\opE}{\mathbb{E}}
\safemath{\opF}{\mathbb{F}}
\safemath{\opG}{\mathbb{G}}
\safemath{\opH}{\mathbb{H}}
\safemath{\opI}{\mathbb{I}}
\safemath{\opJ}{\mathbb{J}}
\safemath{\opK}{\mathbb{K}}
\safemath{\opL}{\mathbb{L}}
\safemath{\opM}{\mathbb{M}}
\safemath{\opN}{\mathbb{N}}
\safemath{\opO}{\mathbb{O}}
\safemath{\opP}{\mathbb{P}}
\safemath{\opQ}{\mathbb{Q}}
\safemath{\opR}{\mathbb{R}}
\safemath{\opS}{\mathbb{S}}
\safemath{\opT}{\mathbb{T}}
\safemath{\opU}{\mathbb{U}}
\safemath{\opV}{\mathbb{V}}
\safemath{\opW}{\mathbb{W}}
\safemath{\opX}{\mathbb{X}}
\safemath{\opY}{\mathbb{Y}}
\safemath{\opZ}{\mathbb{Z}}
\safemath{\opZero}{\mathbb{O}}
\safemath{\identityop}{\opI}

\safemath{\veca}{\bma}
\safemath{\vecb}{\bmb}
\safemath{\vecc}{\bmc}
\safemath{\vecd}{\bmd}
\safemath{\vece}{\bme}
\safemath{\vecf}{\bmf}
\safemath{\vecg}{\bmg}
\safemath{\vech}{\bmh}
\safemath{\veci}{\bmi}
\safemath{\vecj}{\bmj}
\safemath{\veck}{\bmk}
\safemath{\vecl}{\bml}
\safemath{\vecm}{\bmm}
\safemath{\vecn}{\bmn}
\safemath{\veco}{\bmo}
\safemath{\vecp}{\bmp}
\safemath{\vecq}{\bmq}
\safemath{\vecr}{\bmr}
\safemath{\vecs}{\bms}
\safemath{\vect}{\bmt}
\safemath{\vecu}{\bmu}
\safemath{\vecv}{\bmv}
\safemath{\vecw}{\bmw}
\safemath{\vecx}{\bmx}
\safemath{\vecy}{\bmy}
\safemath{\vecz}{\bmz}

\safemath{\veczero}{\bmzero}
\safemath{\vecone}{\bmone}
\safemath{\vecxi}{\bmxi}
\safemath{\veclambda}{\bmlambda}
\safemath{\vecmu}{\bmmu}
\safemath{\vectheta}{\bmtheta}
\safemath{\vecphi}{\bmphi}
\safemath{\vecdelta}{\bmdelta}

\safemath{\matA}{\bA}
\safemath{\matB}{\bB}
\safemath{\matC}{\bC}
\safemath{\matD}{\bD}
\safemath{\matE}{\bE}
\safemath{\matF}{\bF}
\safemath{\matG}{\bG}
\safemath{\matH}{\bH}
\safemath{\matI}{\bI}
\safemath{\matJ}{\bJ}
\safemath{\matK}{\bK}
\safemath{\matL}{\bL}
\safemath{\matM}{\bM}
\safemath{\matN}{\bN}
\safemath{\matO}{\bO}
\safemath{\matP}{\bP}
\safemath{\matQ}{\bQ}
\safemath{\matR}{\bR}
\safemath{\matS}{\bS}
\safemath{\matT}{\bT}
\safemath{\matU}{\bU}
\safemath{\matV}{\bV}
\safemath{\matW}{\bW}
\safemath{\matX}{\bX}
\safemath{\matY}{\bY}
\safemath{\matZ}{\bZ}
\safemath{\matzero}{\bmzero}

\safemath{\matDelta}{\bDelta}
\safemath{\matLambda}{\bLambda}
\safemath{\matPhi}{\bPhi}
\safemath{\matSigma}{\bSigma}
\safemath{\matOmega}{\bOmega}
\safemath{\matTheta}{\bTheta}

\safemath{\matidentity}{\matI}
\safemath{\matone}{\matO}

\safemath{\rnda}{A}
\safemath{\rndb}{B}
\safemath{\rndc}{C}
\safemath{\rndd}{D}
\safemath{\rnde}{E}
\safemath{\rndf}{F}
\safemath{\rndg}{G}
\safemath{\rndh}{H}
\safemath{\rndi}{I}
\safemath{\rndj}{J}
\safemath{\rndk}{K}
\safemath{\rndl}{L}
\safemath{\rndm}{M}
\safemath{\rndn}{N}
\safemath{\rndo}{O}
\safemath{\rndp}{P}
\safemath{\rndq}{Q}
\safemath{\rndr}{R}
\safemath{\rnds}{S}
\safemath{\rndt}{T}
\safemath{\rndu}{U}
\safemath{\rndv}{V}
\safemath{\rndw}{W}
\safemath{\rndx}{X}
\safemath{\rndy}{Y}
\safemath{\rndz}{Z}

\safemath{\rveca}{\bimA}
\safemath{\rvecb}{\bimB}
\safemath{\rvecc}{\bimC}
\safemath{\rvecd}{\bimD}
\safemath{\rvece}{\bimE}
\safemath{\rvecf}{\bimF}
\safemath{\rvecg}{\bimG}
\safemath{\rvech}{\bimH}
\safemath{\rveci}{\bimI}
\safemath{\rvecj}{\bimJ}
\safemath{\rveck}{\bimK}
\safemath{\rvecl}{\bimL}
\safemath{\rvecm}{\bimM}
\safemath{\rvecn}{\bimN}
\safemath{\rveco}{\bomO}
\safemath{\rvecp}{\bimP}
\safemath{\rvecq}{\bimQ}
\safemath{\rvecr}{\bimR}
\safemath{\rvecs}{\bimS}
\safemath{\rvect}{\bimT}
\safemath{\rvecu}{\bimU}
\safemath{\rvecv}{\bimV}
\safemath{\rvecw}{\bimW}
\safemath{\rvecx}{\bimX}
\safemath{\rvecy}{\bimY}
\safemath{\rvecz}{\bimZ}

\safemath{\rvecxi}{\bmxi}
\safemath{\rveclambda}{\bmlambda}
\safemath{\rvecmu}{\bmmu}
\safemath{\rvectheta}{\bmtheta}
\safemath{\rvecphi}{\bmphi}

\safemath{\rmatA}{\bimA}
\safemath{\rmatB}{\bimB}
\safemath{\rmatC}{\bimC}
\safemath{\rmatD}{\bimD}
\safemath{\rmatE}{\bimE}
\safemath{\rmatF}{\bimF}
\safemath{\rmatG}{\bimG}
\safemath{\rmatH}{\bimH}
\safemath{\rmatI}{\bimI}
\safemath{\rmatJ}{\bimJ}
\safemath{\rmatK}{\bimK}
\safemath{\rmatL}{\bimL}
\safemath{\rmatM}{\bimM}
\safemath{\rmatN}{\bimN}
\safemath{\rmatO}{\bimO}
\safemath{\rmatP}{\bimP}
\safemath{\rmatQ}{\bimQ}
\safemath{\rmatR}{\bimR}
\safemath{\rmatS}{\bimS}
\safemath{\rmatT}{\bimT}
\safemath{\rmatU}{\bimU}
\safemath{\rmatV}{\bimV}
\safemath{\rmatW}{\bimW}
\safemath{\rmatX}{\bimX}
\safemath{\rmatY}{\bimY}
\safemath{\rmatZ}{\bimZ}

\safemath{\rmatDelta}{\bimDelta}
\safemath{\rmatLambda}{\bimLambda}
\safemath{\rmatPhi}{\bimPhi}
\safemath{\rmatSigma}{\bimSigma}
\safemath{\rmatOmega}{\bimOmega}
\safemath{\rmatTheta}{\bimTheta}

%% file: macros/standard-macros.tex
\usepackage{amssymb}
\usepackage{amsfonts}
\usepackage{mathrsfs}
\usepackage{xspace}
\usepackage{bm}
\usepackage{fancyref}
\usepackage{textcomp}

\usepackage{multirow}
\usepackage{stmaryrd}

\newenvironment{textbmatrix}{	\setlength{\arraycolsep}{2.5pt}%
								\left[\begin{matrix}}{\end{matrix}\right]%
								\raisebox{0.08ex}{\vphantom{M}}}

\def\be{\begin{equation}}
\def\ee{\end{equation}}
\def\een{\nonumber \end{equation}}
\def\mat{\begin{bmatrix}}
\def\emat{\end{bmatrix}}
\def\btm{\begin{textbmatrix}}
\def\etm{\end{textbmatrix}}

\def\ba#1\ea{\begin{align}#1\end{align}}
\def\bas#1\eas{\begin{align*}#1\end{align*}}
\def\bs#1\es{\begin{split}#1\end{split}}
\def\bg#1\eg{\begin{gather}#1\end{gather}}
\def\bml#1\eml{\begin{multline}#1\end{multline}}
\def\bi#1\ei{\begin{itemize}#1\end{itemize}}

\newcommand{\lefto}{\mathopen{}\left}

\DeclareMathOperator*{\argmin}{arg\;min}		

\newcommand{\vecnorm}[1]{\lefto\lVert#1\right\rVert}		
\newcommand{\herm}[1]{\ensuremath{#1^{H}}} 	
\newcommand{\inv}[1]{\ensuremath{#1^{-1}}} 	

\safemath{\dirac}{\delta}					
\safemath{\krond}{\dirac}					

\safemath{\upto}{\uparrow}
\safemath{\downto}{\downarrow}
\safemath{\iu}{j}							
\safemath{\ev}{\lambda}						
\safemath{\hilseqspace}{l^{2}}				
\newcommand{\banachfunspace}[1]{\setL^{#1}}	
\safemath{\hilfunspace}{\banachfunspace{2}}	

\safemath{\SNR}{\textit{SNR}} 				
\safemath{\PAR}{\textit{PAR}} 				
\safemath{\No}{N_0}							
\safemath{\Es}{E_s}							
\safemath{\Eb}{E_b}							
\safemath{\EbNo}{\frac{\Eb}{\No}}
\safemath{\EsNo}{\frac{\Es}{\No}}

\DeclareMathOperator{\CHop}{\ensuremath{\opH}} 
\safemath{\tvir}{\rndh_{\CHop}}				
\safemath{\tvtf}{\rndl_{\CHop}}				
\safemath{\spf}{\rnds_{\CHop}}				
\safemath{\bff}{H_{\CHop}}					

\safemath{\ircf}{r_{h}}						
\safemath{\tftvcf}{r_{s}}					
\safemath{\tfcf}{r_{l}}						
\safemath{\bfcf}{r_{H}}						

\safemath{\tcorr}{c_h}						
\safemath{\scf}{c_{s}}						
\safemath{\tfcorr}{c_{l}}					
\safemath{\fcorr}{c_{H}}						

\safemath{\mi}{I}							
\safemath{\capacity}{C}						

\safemath{\normal}{\mathcal{N}}			
\safemath{\jpg}{\mathcal{CN}}			
\safemath{\mchain}{\leftrightarrow}		

\safemath{\dB}{\,\mathrm{dB}}
\safemath{\dBm}{\,\mathrm{dBm}}
\safemath{\Hz}{\,\mathrm{Hz}}
\safemath{\kHz}{\,\mathrm{kHz}}
\safemath{\MHz}{\,\mathrm{MHz}}
\safemath{\GHz}{\,\mathrm{GHz}}
\safemath{\s}{\,\mathrm{s}}
\safemath{\ms}{\,\mathrm{ms}}
\safemath{\mus}{\,\mathrm{\text{\textmu}s}}
\safemath{\ns}{\,\mathrm{ns}}
\safemath{\ps}{\,\mathrm{ps}}
\safemath{\meter}{\,\mathrm{m}}
\safemath{\mm}{\,\mathrm{mm}}
\safemath{\cm}{\,\mathrm{cm}}
\safemath{\m}{\,\mathrm{m}}
\safemath{\W}{\,\mathrm{W}}
\safemath{\mW}{\, \mathrm{mW}}
\safemath{\J}{\,\mathrm{J}}
\safemath{\K}{\,\mathrm{K}}
\safemath{\bit}{\,\mathrm{bit}}
\safemath{\nat}{\,\mathrm{nat}}

\safemath{\define}{\triangleq}			

\safemath{\equivalent}{\sim}
\safemath{\distas}{\sim}					
\safemath{\sdiff}{\Delta}				

\safemath{\reals}{\mathbb{R}}
\safemath{\positivereals}{\reals_{+}}
\safemath{\integers}{\mathbb{Z}}
\safemath{\posint}{\integers_{+}}
\safemath{\naturals}{\mathbb{N}}
\safemath{\posnaturals}{\naturals_{+}}
\safemath{\complexset}{\mathbb{C}}
\safemath{\rationals}{\mathbb{Q}}

\newcommand*{\fancyrefapplabelprefix}{app}		
\newcommand*{\fancyrefthmlabelprefix}{thm}		
\newcommand*{\fancyreflemlabelprefix}{lem}		
\newcommand*{\fancyrefcorlabelprefix}{cor}		
\newcommand*{\fancyrefdeflabelprefix}{def}		
\newcommand*{\fancyrefproplabelprefix}{prop}		
\newcommand*{\fancyrefexmpllabelprefix}{exmpl}
\newcommand*{\fancyrefalglabelprefix}{alg}		
\newcommand*{\fancyreftbllabelprefix}{tbl}		

\frefformat{vario}{\fancyrefseclabelprefix}{Sec.~#1}
\frefformat{vario}{\fancyrefthmlabelprefix}{Thm.~#1}
\frefformat{vario}{\fancyreftbllabelprefix}{Tbl.~#1}
\frefformat{vario}{\fancyreflemlabelprefix}{Lem.~#1}
\frefformat{vario}{\fancyrefcorlabelprefix}{Corr.~#1}
\frefformat{vario}{\fancyrefdeflabelprefix}{Def.~#1}
\frefformat{vario}{\fancyreffiglabelprefix}{Fig.~#1}
\frefformat{vario}{\fancyrefapplabelprefix}{App.~#1}
\frefformat{vario}{\fancyrefeqlabelprefix}{(#1)}
\frefformat{vario}{\fancyrefproplabelprefix}{Prop.~#1}
\frefformat{vario}{\fancyrefexmpllabelprefix}{Ex.~#1}
\frefformat{vario}{\fancyrefalglabelprefix}{Alg.~#1}

%% file: macros/defs.tex
\safemath{\dictab}{[\,\dicta\,\,\dictb\,]}

\safemath{\ysig}{\bmy}
\safemath{\ysighat}{\hat{\ysig}}
\safemath{\ysigdim}{M}
\safemath{\xsig}{\bmx}
\safemath{\xsigdim}{N}
\safemath{\nx}{n_x}
\safemath{\zsig}{\bmz}
\safemath{\zsigdim}{\ysigdim}
\safemath{\rsig}{\bmr}
\safemath{\Adict}{\bA}
\safemath{\Adicttilde}{\widetilde{\Adict}}
\safemath{\Adictdim}{\outputdim\times\xsigdim}
\safemath{\avec}{\bma}
\safemath{\avectilde}{\tilde{\avec}}
\safemath{\Bdict}{\bB}
\safemath{\Bdicttilde}{\widetilde{\Bdict}}
\safemath{\Cdict}{\bC}
\safemath{\cvec}{\bmc}
\safemath{\Ddict}{\bD}
\safemath{\Ddictdim}{\ysigdim\times\xsigdim}
\safemath{\dvec}{\bmd}
\safemath{\Ddicttilde}{\widetilde{\bD}}
\safemath{\Bonb}{\bB}
\safemath{\bvec}{\bmb}
\safemath{\Bonbdim}{\ysigdim\times\ysigdim}
\safemath{\noise}{\bmn}
\safemath{\noisedim}{\ysigim}
\safemath{\err}{\bme}
\safemath{\errdim}{\ysigdim}
\safemath{\errset}{\setE}
\safemath{\nerr}{n_e}
\safemath{\delop}{\bP_\errset}
\safemath{\delopc}{\bP_{{\errset}^c}}

\safemath{\cplxi}{\imath}
\safemath{\cplxj}{\jmath}

\safemath{\dict}{\matD}
\safemath{\inputdim}{N}		
\safemath{\outputdim}{M}		
\safemath{\sparsity}{S}	
\safemath{\inputdimA}{{N_a}}	
\safemath{\inputdimB}{{N_b}}	
\safemath{\elemA}{{n_a}}	
\safemath{\elemB}{{n_b}}	
\safemath{\resA}{\matR_a}	
\safemath{\resB}{\matR_b}	
\safemath{\subD}{\matS} 
\safemath{\subA}{\matS_a} 
\safemath{\subB}{\matS_b} 
\safemath{\dicta}{\matA} 	
\safemath{\dictb}{\matB} 	
\safemath{\hollowS}{H}
\safemath{\hollowA}{H_a}
\safemath{\hollowB}{H_b}
\safemath{\cross}{Z}
\safemath{\coh}{\mu_d}			
\safemath{\coha}{\mu_a}			
\safemath{\cohb}{\mu_b}			
\safemath{\mubs}{\nu}	
\safemath{\cohm}{\mu_m} 
\safemath{\dictset}{\setD}	
\safemath{\dictsetp}{\dictset(\coh,\coha,\cohb)}	
\safemath{\dictsetgen}{\dictset_\text{gen}}
\safemath{\dictsetgenp}{\dictsetgen(\coh)}
\safemath{\dictsetonb}{\dictset_\text{onb}}
\safemath{\dictsetonbp}{\dictsetonb(\coh)}

\safemath{\leftside}{U}
\safemath{\rightsideA}{R_a}
\safemath{\rightsideB}{R_b}

\safemath{\indexS}{\setI_S} 

\safemath{\na}{n_a}			
\safemath{\nb}{n_b}			
\safemath{\coeffa}{p_i}	
\safemath{\coeffb}{q_j}	
\safemath{\seta}{\setP}		
\safemath{\setb}{\setQ}     
\safemath{\setw}{\setW}	
\safemath{\setz}{\setZ}	
\safemath{\cola}{\veca}		
\safemath{\colb}{\vecb}		
\safemath{\cold}{\vecd}		
\safemath{\inputvec}{\vecx} 	
\safemath{\error}{\vece}	
\safemath{\noiseout}{\vecz} 	
\safemath{\inputvecel}{x}
\safemath{\inputveca}{\vecx_a}
\safemath{\inputvecb}{\vecx_b}
\safemath{\outputvec}{\vecy}	
\safemath{\lambdamin}{\lambda_{\mathrm{min}}}

\safemath{\elltwo}{\ell_2}
\safemath{\ellone}{\ell_1}
\safemath{\ellzero}{\ell_0}
\safemath{\ellinf}{\ell_\infty}
\safemath{\ellinftilde}{\ell_{\widetilde\infty}}
\safemath{\licard}{Z(\coh,\coha,\cohb)}
\safemath{\xsol}{\hat{x}}
\safemath{\xbord}{x_b}		
\safemath{\xstat}{x_s}		
\safemath{\xstatLone}{\tilde{x}_s}
\safemath{\order}{\mathcal{O}} 
\safemath{\scales}{\Theta} 
\safemath{\ones}{\mathbf{1}} 
\safemath{\zeroes}{\mathbf{0}} 
\safemath{\thlone}{\kappa(\coh,\cohb)} 
\safemath{\constoneA}{\delta} 
\safemath{\constoneB}{\epsilon} 
\safemath{\nlarge}{L}				   
\safemath{\sumlarge}{S_\nlarge}
\safemath{\maxlarger}{P_\nlarge}	   
\safemath{\Pzero}{\textrm{P0}}	
\safemath{\Pone}{\textrm{P1}}
\safemath{\vecfir}{\vecw}			 
\safemath{\vecsec}{\vecz}
\safemath{\elvecfir}{w}              
\safemath{\elvecsec}{z}				 
\safemath{\nlargefir}{n}
\safemath{\normout}{\gamma}
\safemath{\auxfun}{h}
\safemath{\supp}{\textrm{supp}}

\safemath{\indexa}{\ell}
\safemath{\indexb}{r}
\safemath{\indexc}{i}
\safemath{\indexd}{j}

\safemath{\project}{P}

\safemath{\firstslotset}{\setU_1}  
\safemath{\secondslotset}{\setU_2} 
\safemath{\randomset}{\setS} 
\newcommand{\electionone}[1]{#1} 
\newcommand{\electiontwo}[1]{#1'}